\documentclass[oldversion]{aa} 
\usepackage{graphicx}
\usepackage{aalongtable}
\usepackage{txfonts}
\usepackage{rotating}
\usepackage{lscape}
\newcommand{\gsim}{\;\lower.6ex\hbox{$\sim$}\kern-7.75pt\raise.65ex\hbox{$>$}\;}
\newcommand{\lsim}{\;\lower.6ex\hbox{$\sim$}\kern-7.75pt\raise.65ex\hbox{$<$}\;}

\begin{document}
\title{Metallicity of the globular cluster NGC~6388 from high
resolution spectra of more than 160 giant stars\thanks{Based on observations collected at 
ESO telescopes under programmes 073.D-0211, and 073.D-0760,  381.D-0329,  
095.D-0834.}
\thanks{Full Tables 2 and 3 are only available at the CDS via anonymous ftp to cdsarc.u-strasbg.fr (130.79.128.5) or via http://cdsarc. u-strasbg.fr/viz-bin/cat/J/A+A/??/??}
 }

\author{
Eugenio Carretta\inst{1}
\and
Angela Bragaglia\inst{1}
}

\authorrunning{Carretta and Bragaglia}
\titlerunning{Metallicity of NGC~6388}

\offprints{E. Carretta, eugenio.carretta@inaf.it}

\institute{
INAF-Osservatorio di Astrofisica e Scienza dello Spazio di Bologna, via Gobetti
 93/3, I-40129 Bologna, Italy \\
 {\tt e-mail: eugenio.carretta@inaf.it, angela.bragaglia@inaf.it}}

\date{}

\abstract{NGC~6388 is one of the most massive Galactic globular clusters (GC) and it
is an old, metal-rich, Galactic bulge cluster. By exploiting previous
spectroscopic observations we were able to bypass the uncertainties in
membership related to the strong field stars contamination. We present the
abundance analysis of 12 new giant stars with UVES spectra and 150 giants
with GIRAFFE spectra acquired at the ESO-VLT. We derived radial velocities,
atmospheric parameters and iron abundances for all stars. When combined to
previous data, we obtain a grand total of 185 stars homogeneously analysed in
NGC~6388 from high-resolution spectroscopy. The average radial velocity of
the 185 stars is 81.2$\pm0.7$, rms 9.4 km~s$^{-1}$. We obtain an average
metallicity [Fe/H]$=-0.480$ dex, $rms = 0.045$ dex (35 stars) and
[Fe/H]$=-0.488$ dex,  $rms = 0.040$ dex (150 stars) from the UVES and GIRAFFE
samples, respectively. Comparing these values to internal errors in abundance,
we exclude the presence of a significant intrinsic metallicity spread within the
cluster. Since about a third of giants in NGC 6388 is claimed to belong to the
``anomalous red giants" in the HST pseudo-colour map defining the so-called
type-II GCs, we conclude that either enhanced metallicity is not a necessary
requisite to explain this classification (as also suggested by the null iron
spread for NGC~362)  or NGC~6388 is not a type-II globular cluster.
}

\maketitle

\section{Introduction}

The metal-rich and massive bulge globular cluster (GC) NGC~6388 
([Fe/H]=-0.44\footnote{We adopt the usual spectroscopic notation, $i.e.$  [X]=
log(X)$_{\rm star} -$ log(X)$_\odot$ for any abundance quantity X, and  log
$\epsilon$(X) = log (N$_{\rm X}$/N$_{\rm H}$) + 12.0 for absolute number density
abundances.}, Carretta et al. 2007a; $M_V$=-9.41 mag, Harris 1996, 2010 web
edition) certainly
deserves attention due to its many peculiar features: {\em a)} beside the red
horizontal branch (HB) typical of old, metal-rich GCs, NGC~6388 shows an
extended blue HB, i.e., we see the second parameter at work within the same
cluster. The He enrichment in the second generation stars (as suggested by
D'Antona \& Caloi 2004; see also Gratton et al. 2010), is likely the
explanation, as shown by NGC~6388 participating to the strong correlation
between extension of the Na-O anti-correlation and maximum temperature on the HB
(Carretta et al. 2007b); {\em b)} NGC~6388 is a local counterpart of old,
metal-rich populations found in distant elliptical galaxies, and its relevant
population of hot HB stars is a likely contributor to the UV-upturn phenomenon
(e.g. Yi et al. 1998); {\em c)} NGC~6388 is a pivotal cluster to probe the
existence of intermediate-mass black holes (IMBH $10^3-10^4 M_\odot$) whose
evidence in this GC is controversial (Lanzoni et al. 2013, hereinafter L13;
Lutzgendorf et al. 2015); {\em d)} NGC~6388 has been classified, on the basis of
HST photometry pseudo-colours (the so-called ``chromosome map", ChM,  Milone et
al. 2017) as an example of a sub-class dubbed type-II GCs, i.e., GCs showing
not only a separation between putative first and second populations, but also an
additional ``anomalous" population. According to Milone et al. (2017), 
a type-II GC is characterised by at least
one of the following features: presence of separate subgiant and red giant
branches (SGB, RGB), variation in C+N+O content (and/or age),  enhanced heavy
elements abundances (iron and neutron-capture elements).

In our project on NGC~6388 we concentrate on the last point,  spectroscopically
studying a large sample of cluster stars. We successfully bypassed the problem
of the heavy field contamination affecting this GC by exploiting a large mass of
spectroscopic data acquired for various purposes and contained in the ESO
archive. By forecasting high probability members from their radial velocity, we
were able to derive an homogeneous chemical characterisation of a large sample
of cluster stars using high resolution spectroscopy of individual targets. We
tackle the problem of a possible metallicity dispersion, which seems to be a
common requisite for so called type-II GCs. We present here  the full set of
atmospheric parameters (temperature, gravity, and metallicity) while abundances
of light and heavy elements will be presented in a forthcoming paper.

NGC~6388 is generally classified as a bulge GC, not only from its current
location in the Galaxy (Galactocentric distance 3.1 kpc, 1.2 kpc below the
Galactic plane, Harris 1996, 2010 edition), but also because it follows an
highly bound orbit, with apocenter at less than 3.5 kpc (e.g. Massari et al. 2019).
It is an old GC (Ortolani et al. 1995) and its high metallicity places the
cluster among a near uniform old age group in the age-metallicity relation
(AMR). Based on the location on the AMR and in the integral of motion space,
Forbes (2020) assigns NGC~6388 to the in situ group of Milky Way GCs,
confirming the classification by Massari et al. (2019; but see Myeong et al.
2019 and Minelli et al. 2021 for a different view). 
All this confirms the interest of the cluster, which merits in-depth study, both
of its chemical and astrometric properties.

The existence of a possible metallicity spread in NGC~6388
is controversial. From an apparent, qualitative resemblance of the 
colour-magnitude diagram (CMD) with that
of $\omega$~Cen, Piotto et al. (1997) raised the suspicion of a metallicity
spread to explain both the spread in colour of the RGB and the peculiar HB of
NGC~6388. However, Raimondo et al. (2002) found that a large spread in metal
abundance could not be  reconciled with the CMD features.
From Washington photometry, Hughes et al. (2007) were able to put an upper limit
to the metallicity spread, estimated to be less than 0.2 dex, supporting the
abundance analysis of only eight cool giants by Wallerstein et al. (2007), who
derived a root mean square (r.m.s.) scatter of 0.1 dex.

We were the first to obtain spectra of many stars in NGC~6388, using FLAMES
(Carretta et al. 2007a, 2009b), but, before Gaia and other spectroscopic surveys
on this cluster, the pointing of potential targets was forcefully a shot in the
dark. In fact, NGC~6388 is strongly contaminated by disk and bulge stars. While
use of precise astrometry  could have helped the target selection even in this
very crowded and difficult field, the spectra  were acquired long before Gaia
Data Release 2 (which went public on April 2018) and we could measure Fe, O, and
Na only in 32 member stars, almost the lowest number in our FLAMES survey of GCs
(Carretta et al. 2006, 2009a,b, 2010a). Clearly, this needed to be corroborated
by more robust statistics. 

We then decided to take advantage of the huge potential of the ESO archive.
First results based only on UVES spectra of 24 new member stars were presented
in Carretta and Bragaglia (2018). We measured Fe and 15 other elements, finding
no spread in metallicity. We also confirmed the normal light elements
anti-correlations found in all GCs, with clearly defined groups of primordial,
intermediate, and extreme composition stars (P, I, and E as defined in Carretta
et al. 2009b). 
However, if part of the cluster population is distinct by some difference in
abundance, the fraction can be even small. For instance, in NGC~362 Carretta et al. (2013)
found a secondary giant branch containing only about 6\% of RGB stars, more
Ba-rich of the bulk of normal RGB giants. 

Hence, it is important to analyse samples as large as possible.
L13 presented metallicities for about 280 NGC~6388 giants, partly from direct analysis of
iron equivalent widths (EWs) and partly from the infrared calcium triplet (CaT) method.
They did not found significant metallicity spread, but the detailed abundance
analysis was not published.
Recently, the discussion on a possible spread in metal abundance for NGC~6388 was
resumed following the studies by M\'esz\'aros et al. (2020, hereinafter M20)
from infrared APOGEE spectra and Husser et al. (2020, from now H20) from low
resolution MUSE spectra.

Here we present the whole sample of archive and  new data acquired to give  a
clear-cut answer to the issue of the intrinsic metallicity spread (or the lack of
it) in NGC~6388, exploiting a large sample of stars with high resolution spectra
analysed in the most homogeneous way. We then provide abundances of iron for
individual stars as well as the atmospheric parameters derived on the same scale
as in our previous works on more than 25 GCs. These values will be used in a forthcoming
paper devoted to fully explore the chemical characterisation of multiple
population in NGC~6388, their link with the dynamics, the clues from chemistry
on the origin of this GCs, and other issues.
The present paper is organised as follows: the data sets are described in
Sect.~2 and the analysis in Sect.~3. Results on the metallicity distribution and
comparison with literature data are presented in Sect.~4; a discussion of
NGC~6388 in the context of type-II GCs is in Sect.~5; finally, a summary is 
given in Sect.~6.

\section{Sample selection, observations and previous data}

\begin{table*}
\centering
\caption{Proprietary and archive data used in this paper.}
\begin{tabular}{lllrl}
\hline
 Programme & sample & Obs date & Obs time & setups \\
 \hline
073.D-0211 (PI Carretta) &C09 & May to July 2004  & 19.6 hr & HR11, HR13, U580 \\
099.D-0047 (PI Carretta) &us  & April to August 2017 & 11.9 hr & HR13, HR21, U580 \\ 
073.D-0760 (PI Catelan)  &CAT & July 2004         & 3 hr    & HR13, U580 \\
381.D-0329 (PI Lanzoni)  &L13 & June to July 2008 & 3 hr    & HR21, U580 \\
095.D-0834 (PI Henault-Brunet) &H-B &  June 2015  & 3.8 hr  & HR13, U580 \\
\hline
\end{tabular}
\begin{list}{}{}
\item[Wavelength coverage -]
HR11: 5597-5840;
HR13: 6120-6405; 
HR21: 8484-9001;
U580: 4800-6800 \AA. The GIRAFFE setups prior and later than February 2015 have slightly different resolutions, see ESO web pages. 
\end{list}
\label{t:log}
\end{table*}

The choice of targets in our latest observing programme (099.D-0047) was helped
by the existing data sets available in the ESO archive, providing membership.
This was our starting point, a key step to avoid wasting telescope time on
contaminating field stars whose density is very high  toward NGC~6388 and which
plagued our original programme (073.D-0211).

To ensure an optimal return from the observing time spent on targets we
considered three main data sets present in the ESO archive of advanced data
products (see Table~\ref{t:log}). All spectra considered were acquired with
FLAMES (Pasquini et al. 2000). There are 539 GIRAFFE spectra taken with the
high-resolution setup HR21 (R=18000, wavelength coverage $\sim 8480- 9000$~\AA)
for 398 stars from L13 (ESO programme 381.D-0329). From ESO programme 95.D-0834
(P.I. Henault-Brunet, hereinafter H-B sample) we retrieved 1112 GIRAFFE HR13
spectra (R=24000, spectral range $\sim 6115-6400$~\AA) for 113 stars. Finally,
from the programme 073.D-0760 (P.I. Catelan; hereinafter CAT sample) we found
113 GIRAFFE HR13 spectra. 

To select member stars we adopted the range in radial velocity (RV) as given in
L13, whose large sample was designed to study the velocity dispersion in
NGC~6388. Restricting the RVs to the interval 50-110 km~s$^{-1}$ (as in L13) we
obtained 276, 65, and 53 candidate member stars from L13, H-B, and CAT,
respectively. This was the original pool from which to cull our final targets.

A posteriori, we cross-matched our total 185 targets with the table of membership probability  by
Vasiliev \& Baumgardt (2021), based on Gaia Early Data Release 3 results (Gaia
Collaboration et al. 2021), We found 176 stars in common. For 90\% of them the
astrometric membership probability is in agreement with the one based on RV: 152
stars out of 176 have probability larger than 0.9, and 7 more have probability
between 0.5 and 0.9. Given the different methods and the still not optimal
performance of Gaia in dense fields, we deem this a successful comparison.

The criteria for configuring the pointings were driven by the available archive
material as well as by our main purposes: to derive a clearcut answer on the
intrinsic metallicity distribution in NGC~6388 and to obtain a detailed chemical
characterisation of the multiple stellar populations hosted in this GC. 
Our observing strategy was then set as follow.
For the stars observed only with HR21 we can derive Al abundances, so we
need to complement them with HR13 spectra to derive also O, Na, Mg, beside
metallicity and atmospheric parameters in a homogeneous way.
For stars with HR13 spectra only, we want to add HR21 spectra to derive
also Al abundances to obtain the complete set of proton-capture elements
involved in (anti-)correlations defining multiple stellar populations in GCs.

Taking the sample by L13 as main source of targets, we counter-identified the
276 member stars with HR21 spectra with the the photometry obtained with  
the Wide Field Imager (WFI) at the ESO 2.2m telescope, used for our
previous analysis and described in Carretta et al. (2007a). We selected only
stars with $V<16.25$ to avoid excessively long exposure times. To these we added
stars from the H-B sample with low S/N in their HR13 spectra. Ten one-hour
exposures were obtained in Service Mode by ESO personnel in 2017 (see
Table~\ref{t:log}).

As a complementary data set we considered stars that had HR13 observations but
lacked HR21 spectra to define a second configuration for the observations with
the HR21 setup. Two exposures (3160 sec each) were obtained on 24 and 28 July
2017.

Finally, FLAMES/UVES fibres were put on the brightest giants  to obtain the full
pattern of the chemical composition, from the lightest to the heaviest
(neutron-capture) species (using the 580 setup). During the previously described
exposures we then observed 12 member stars at R$\sim 47000$ and with
wavelength coverage from about 4800 to 6800~\AA.

The observations with the setup HR13 were combined, if feasible, with archival
data or spectra for the limited sample of stars used in Carretta et al. (2009b).
From the latter, we only considered the HR13 setup, to achieve the maximum
possible degree of homogeneity. Most of the the Fe~{\sc i}  and Fe~{\sc ii}
lines to derive the metal abundance fall in the spectral range of HR13 spectra,
as well as transitions of the O, Mg, Si, Ca, and Sc species. Neglecting the HR11
setup, where the Na doublet at 5682-88 \AA\ lies, has no consequences, as
the weaker lines of Na at 6154-60~\AA\ are strong enough to secure precise 
abundances of this element at the high metallicity of NGC~6388.

\begin{table*}
\centering
\caption{Names, original sample, coordinates, magnitudes, and radial velocities
of program stars. (The complete Table is available at CDS, Strasbourg).}
\begin{tabular}{llrlllllcr}
\hline
 star	 &  sample  &S/N &  RA         &    DEC     &  V(wfi)  &B(wfi) &K(2ma) & wfi & RV \\
\hline
 n63a	 &  UVES    & 30 &  264.080826 & -44.733165 &	14.968 &16.511 &10.078 &  1 &  75.54 \\   
 n63b	 &  UVES    & 30 &  264.089261 & -44.728379 &	15.18  &       &9.733  &  0 &  62.66 \\
 n63c	 &  UVES    & 40 &  264.096031 & -44.745935 &	14.99  &16.995 &9.618  &  1 &  85.77 \\   
 n63d	 &  UVES    & 50 &  264.073095 & -44.744597 &	14.804 &16.616 &9.694  &  1 &  88.10 \\   
 n63e	 &  UVES    & 50 &  264.094481 & -44.75657  &	15.188 &17.097 &10.394 &  1 & 105.47 \\   
 n63f	 &  UVES    & 60 &  264.045909 & -44.750423 &	15.14  &17.009 &10.381 &  1 &  81.56 \\   
 n63g	 &  UVES    & 40 &  264.143672 & -44.71806  &	15.252 &17.384 &9.57   &  1 &  72.48 \\   
 n63h	 &  UVES    & 50 &  264.129650 & -44.72916  &	15.071 &17.04  &9.954  &  1 &  79.72 \\   
 n63i	 &  UVES    & 65 &  264.069704 & -44.706032 &	15.161 &17.095 &10.218 &  1 &  71.44 \\   
 n63l	 &  UVES    & 45 &  264.047492 & -44.766651 &	15.233 &17.033 &10.707 &  1 &  76.80 \\   
 n63m	 &  UVES    & 55 &  264.215366 & -44.728741 &	15.18  &       &10.115 &  0 &  80.39 \\
 n63n	 &  UVES    & 30 &  264.033217 & -44.736818 &	14.757 &16.678 &9.731  &  1 &  83.65 \\   
 l63p001 &  H-B     & 60 &  263.972502 & -44.765736 &	16.832 &18.329 &13.366 &  1 &  89.83 \\   
 l63p002 &  H-B     & 49 &  263.863453 & -44.752747 &	17.064 &18.469 &13.745 &  1 &  96.12 \\   
 l63p003 &  C09     & 90 &  263.952259 & -44.750931 &	17.701 &19.061 &14.457 &  1 &  87.29 \\   
 l63p004 &  H-B     & 58 &  263.940325 & -44.717125 &	16.701 &18.139 &13.21  &  1 &  83.73 \\   
 l63p006 &  us      & 20 &  263.962607 & -44.686058 &	17.717 &19.106 &14.212 &  1 &  74.61 \\      
 l63p007 &  H-B     &134 &  263.958385 & -44.677128 &	15.279 &17.180 &10.68  &  1 &  92.17 \\   
 l63p008 &  H-B     & 52 &  263.869447 & -44.646568 &	16.701 &18.291 &12.676 &  1 &  68.01 \\      
 l63p010 &  us      & 45 &  264.168471 & -44.787724 &	17.62  &18.981 &       &  1 &  79.07 \\   
 l63p011 &  C09     &120 &  264.016210 & -44.786659 &	16.574 &18.029 &12.934 &  1 &  79.13 \\   
 l63p012 &  H-B     & 70 &  264.024788 & -44.783100 &	16.48  &17.955 &12.756 &  1 &  76.54 \\   
 l63p013 &  H-B     & 53 &  264.088182 & -44.782795 &	17.053 &18.429 &13.622 &  1 &  67.41 \\   
 l63p014 &  C09     &120 &  264.090260 & -44.774242 &	17.556 &18.876 &14.408 &  1 &  74.66 \\   
 l63p015 &  CAT+H-B &112 &  264.124013 & -44.766060 &	15.985 &17.613 &11.931 &  1 &  82.63 \\      
 l63p017 &  H-B     & 55 &  264.163561 & -44.733860 &	16.944 &18.423 &13.195 &  1 &  83.42 \\   
 l63p018 &  C09     &130 &  264.163546 & -44.733825 &	16.582 &18.143 &       &  1 &  88.95 \\   
\hline
\end{tabular}
\begin{list}{}{}
\item[Note 1:] star n63n is star A04 in Carretta and Bragaglia (2018)
\item[Note 2:] sample: us=this work, new observations, H-B=Henault-Brunet, C09=Carretta et al. (2009b)
CAT=Catelan, L13=Lanzoni et al. (2013)
\item[Note 3:] wfi=1 (0) if the star is present (absent) in the WFI catalogue (see text).
\end{list}
\label{t:coomag}
\end{table*}

In Table~\ref{t:coomag} we list the relevant information to identify all stars
in our sample. Since the naming convention is different in each sub-sample (L13,
CAT, H-B) we used our unique code (column 1) to identify each star. The
coordinates RA and DEC easily allow the cross-identification among all the
sub-samples. The source for the spectra used for each stars and the estimated
S/N ratio are listed in columns 2 and 3. We also provide in Table~\ref{t:coomag}
optical magnitudes obtained with the WFI (see Carretta et al. 2007a). Near
infrared $K$ magnitudes are from 2MASS (Skrutskie et al. 2006). A $V$ magnitude
derived from a $V$ versus $K$ calibration based on all stars with both
magnitudes available was assigned to stars not present in the WFI photometry
(flag=0 in the last column of Table~\ref{t:coomag}).

The total sample of the present work includes 150 stars with GIRAFFE HR13
spectra and 12 with UVES spectra. To it, we can also add the 24 stars with UVES
spectra  homogeneously analysed in Carretta and Bragaglia (2018). Our final
sample in NGC~6388 is then composed by 185 giants (star n63n in the present
study is also star A04 in Carretta and Bragaglia 2018). 

\begin{figure}
\centering
\includegraphics[scale=0.40]{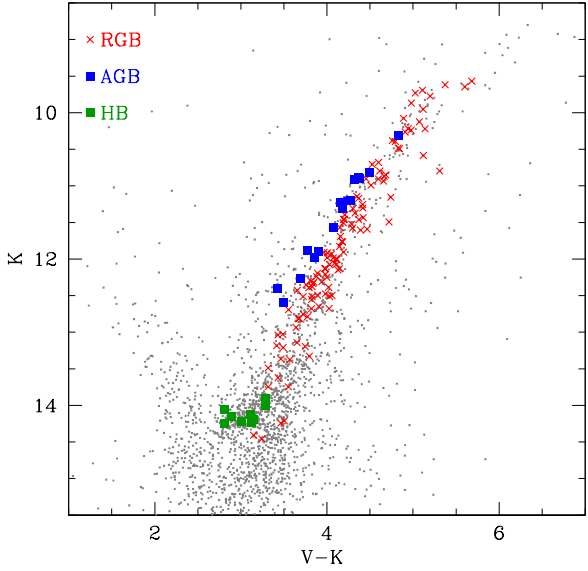}
\caption{$K,V-K$ colour magnitude diagram of NGC~6388 from WFI and 2MASS
photometry (grey dots). Stars observed with spectroscopy in the present work are
labelled and indicated with larger symbols: red crosses, blue squares, and green
squares for RGB, AGB, and RHB stars, respectively.}
\label{f:cmdVKK}
\end{figure}

Differently from other GCs studied in our FLAMES survey, where we selected stars
lying near the RGB ridge line, in NGC~6388 we followed selection criteria
based on RV, so that the final sample contains small sub-samples of
stars in other evolutionary stages. From the position in the CMDs we identified
17 stars on the asymptotic giant branch (AGB) and 10 stars on the red horizontal
branch (RHB). Together with the bulk of RGB stars in our sample, they are plotted
on the $K,V-K$ colour magnitude diagram in Fig.~\ref{f:cmdVKK}. 

\begin{figure}
\centering
\includegraphics[scale=0.40]{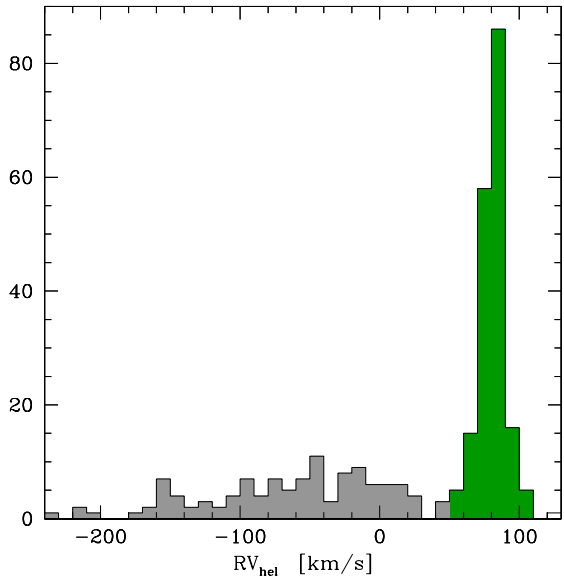}
\caption{Histogram of the RV values, from this paper, Carretta et al. (2007a,
2009b), and Carretta and Bragaglia (2018). We show in green the candidate
members and in grey the non members (one star, at large positive RV, is not
shown).}
\label{f:rv}
\end{figure}

\begin{figure}
\centering
\includegraphics[scale=0.40]{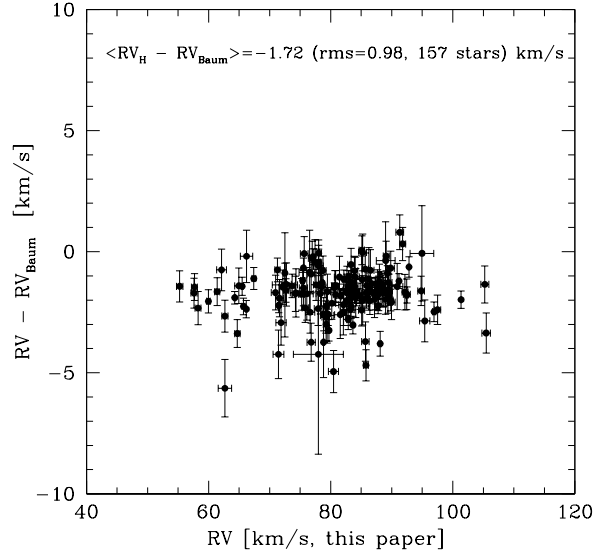}
\caption{Difference in RV between our sample and the data at the Baumgardt
website. Two outliers, with large $\Delta$RV values, are omitted (see text) in the
plot.}
\label{f:drv}
\end{figure}

The histogram of the RV (measured using the {\sc iraf} task {\em rvidlines}) of
these 185 stars is shown in Fig.~\ref{f:rv}, together with the distribution of
non members from the archive data and our original observations. The
average radial velocity of the 185 stars considered cluster members is
81.2$\pm0.7$, rms 9.4 km~s$^{-1}$. Finally, we
cross-matched our targets with the table of individual stellar radial velocities
of stars in the fields of globular clusters available at Holger Baumgardt's
web page\footnote{\tt https://people.smp.uq.edu.au/HolgerBaumgardt/globular/},
finding 159 stars in common. A plot of the differences in RV is shown in
Fig.~\ref{f:drv}, where two stars with differences larger than 10 km~s$^{-1}$
are omitted. Without them, the average offset is -1.72 (rms 0.98) km~s$^{-1}$.
Star l63p154, in particular, shows a difference $+68.11$ km~s$^{-1}$ and a
possible explanation is that the star is a binary. Our observations, taken in a
short time interval, do not display significant changes around +85 km~s$^{-1}$;
however, the value listed at Baumgardt's web page is 17.46 km~s$^{-1}$ while L13
have 2.8  km~s$^{-1}$. A more detailed comparison is outside the main goal of
our paper and we do not proceed further.

\section{Abundance analysis and error budget}

The derivation of metallicity and of atmospheric parameters for all
stars in our sample are based on $EW$s of neutral and singly
ionised iron lines. We measured $EW$s with the package ROSA (Gratton 1988) as
described in Bragaglia et al. (2001). The line list is the same used in all the
papers of our FLAMES survey to study the Na-O anti-correlation in GCs (presented
in Carretta et al. 2006, 2010a). The solar reference abundances are from Gratton
et al. (2003). We recall here that the solar value for iron (the only
element discussed in the present paper) is log $\epsilon$(Fe)=7.54 for Fe~{\sc i} and 7.49 for
Fe~{\sc ii}. For comparison, other reference solar values for iron commonly used
are, for instance, 7.50 (Grevesse \& Sauval 1998), 7.45 (Grevesse et al. 2007),
and 7.50 (Asplund et al. 2009).

The $EW$s measured on the GIRAFFE HR13 spectra were converted to the scale of
$EW$s measured on high resolution UVES spectra using lines measured on stars
observed with both spectrographs (four, five, and four stars in our new
observations and from CAT and H-B samples, respectively). UVES spectra are taken
from Carretta et al. (2007a), Carretta and Bragaglia (2018) and from the present
work.

The procedure to derive the atmospheric parameters is the same adopted for the
other works in our FLAMES survey, so that the derived abundances are on a
strictly homogeneous scale.
Effective temperatures (T$_{\rm eff}$) were obtained from a calibration between
magnitudes and values of T$_{\rm eff}$ from de-reddened $V-K$ colours. The latter
temperatures were derived from the Alonso et al. (1999, 2001) relation. For
NGC~6388 we used $K$ magnitudes. that are less affected by reddening than
$V-K$ colours. The rationale for this approach, aimed at reducing the star-to-star
errors, is discussed in detail in Gratton et al. (2007) for NGC~6441, another
bulge cluster heavily affected by extinction.

Surface gravities were obtained from the above temperatures, adopting distance
modulus $(m-M)_V=16.14$ and reddening $E(B-V)=0.37$ from Harris (1996, 2010 on
line edition), bolometric corrections from Alonso et al. (1999), masses of 
0.90 $M_\odot$ and $M_{\rm bol,\odot}=4.75$. The relations $E(V-K)=2.75E(B-V)$,
$A_V=3.1E(B-V)$, and $A_K=0.353E(B-V)$ are taken from Cardelli et al. (1989).

We minimised the slope of the relation between abundances of Fe~{\sc i} and
expected line strength (see Magain 1984) to derive the values of the
microturbulent velocity $v_t$.
From the Kurucz (1993) grid we selected the model whose abundance was equal to
the average abundances from Fe~{\sc i} lines.
Adopted atmospheric parameters and derived abundances of Fe are listed in
Table~\ref{t:atmparUNIFIN} (only an excerpt is shown, the entire Table will be
available on line at CDS). The listed rms scatter is the line-to-line scatter.
On average, we measured 100 Fe~{\sc i} and 13 Fe~{\sc ii} lines from the UVES
spectra and 20 Fe~{\sc i} and 3 Fe~{\sc ii} from the GIRAFFE HR13 spectra.

\setcounter{table}{2}
\begin{table*}
\centering
\caption[]{Atmospheric parameters and derived iron abundances in
NGC~6388 (The complete Table is only available on line at CDS, Strasbourg).}
\begin{tabular}{rccccrcccrrcc}
\hline
Star   &  $T_{\rm eff}$ & $\log$ $g$ & [A/H]  &$v_t$	     & nr & [Fe/H]{\sc i} & $rms$ & nr & [Fe/H{\sc ii} & $rms$ \\
       &     (K)	&  (dex)     & (dex)  &(km s$^{-1}$) &    & (dex)	  &	  &    & (dex)         &       \\
\hline
n63a     & 3772 & 0.80 & $-$0.48 & 1.30 & 125 & $-$0.476 & 0.214 & 11  & $-$0.492 & 0.194  \\ 
n63b     & 3727 & 0.61 & $-$0.51 & 0.27 & 73  & $-$0.509 & 0.314 & 5   & $-$0.483 & 0.199  \\ 
n63c     & 3714 & 0.58 & $-$0.54 & 1.35 & 77  & $-$0.552 & 0.220 & 12  & $-$0.569 & 0.262  \\ 
n63d     & 3722 & 0.63 & $-$0.57 & 1.17 & 83  & $-$0.573 & 0.177 & 18  & $-$0.543 & 0.239  \\ 
n63e     & 3820 & 0.93 & $-$0.59 & 1.78 & 122 & $-$0.494 & 0.151 & 18  & $-$0.454 & 0.200  \\ 
n63f     & 3818 & 0.93 & $-$0.54 & 1.54 & 101 & $-$0.543 & 0.130 & 14  & $-$0.484 & 0.152  \\ 
n63g     & 3708 & 0.52 & $-$0.44 & 0.76 & 73  & $-$0.436 & 0.216 & 10  & $-$0.487 & 0.290  \\ 
n63h     & 3755 & 0.73 & $-$0.54 & 1.61 & 100 & $-$0.536 & 0.184 & 14  & $-$0.520 & 0.234  \\ 
n63i     & 3793 & 0.85 & $-$0.53 & 1.63 & 113 & $-$0.528 & 0.142 & 16  & $-$0.498 & 0.167  \\ 
n63l     & 3874 & 1.07 & $-$0.49 & 1.66 & 126 & $-$0.489 & 0.166 & 15  & $-$0.447 & 0.145  \\ 
n63m     & 3777 & 0.80 & $-$0.47 & 1.46 & 106 & $-$0.468 & 0.186 & 13  & $-$0.481 & 0.206  \\ 
n63n     & 3727 & 0.65 & $-$0.50 & 1.69 & 100 & $-$0.499 & 0.173 & 10  & $-$0.432 & 0.137  \\ 
l63p001  & 4576 & 2.23 & $-$0.49 & 1.63 & 21  & $-$0.488 & 0.106 & 3   & $-$0.524 & 0.077  \\ 
l63p002  & 4712 & 2.40 & $-$0.51 & 1.53 & 24  & $-$0.513 & 0.121 & 3   & $-$0.391 & 0.201  \\ 
l63p003  & 4991 & 2.69 & $-$0.41 & 1.82 & 16  & $-$0.413 & 0.135 & 3   & $-$0.474 & 0.116  \\ 
l63p004  & 4522 & 2.16 & $-$0.50 & 1.31 & 15  & $-$0.504 & 0.094 & 3   & $-$0.435 & 0.104  \\ 
l63p006  & 4892 & 2.56 & $-$0.45 & 2.23 & 19  & $-$0.449 & 0.105 & 4   & $-$0.468 & 0.119  \\ 
\hline
\end{tabular}
\label{t:atmparUNIFIN}
\end{table*}

We estimated star-to-star errors due to uncertainties in the adopted atmospheric
parameters and in $EW$ measurement using our usual procedure, amply described in
Carretta et al. (2009a) for UVES and Carretta et al. (2009b) for GIRAFFE. The
results for the present work are tabulated in Table~\ref{t:sensitivityuUNIFIN}
and Table~\ref{t:sensitivitymUNIFIN} for errors in iron abundances derived from
UVES and GIRAFFE spectra, respectively.

In the main body of these tables we list the sensitivities of abundance ratios
[Fe/H] to changes in the atmospheric parameters, obtained by changing each of
the parameters at the time for all stars, then taking the average. The amount of
the variation is listed in the first row of each table. In the second and third
row, the star-to-star (internal) errors and the systematic errors in each
parameter are listed. In the second column we report the average number of lines
used. Finally, in the last two columns, we show the total internal and
systematic errors, derived by summing in quadrature the contributions of
individual error sources.

\begin{table*}
\centering
\caption[]{Sensitivities of abundance ratios to variations in the atmospheric
parameters and to errors in the equivalent widths, and errors in abundances for
stars of NGC~6388 observed with UVES.}
\begin{tabular}{lrrrrrrrr}
\hline
Element     & Average   & T$_{\rm eff}$ & $\log g$ & [A/H]   & $v_t$    & EWs     & Total   & Total      \\
            & n. lines  &      (K)      &  (dex)   & (dex)   &kms$^{-1}$& (dex)   &Internal & Systematic \\
\hline        
Variation&              &  50           &   0.20   &  0.10   &  0.10    &         &         &            \\
Internal &              &   6           &   0.04   &  0.04   &  0.12    & 0.02   &         &            \\
Systematic&             &  17           &   0.06   &  0.02   &  0.04    &         &         &            \\
\hline
$[$Fe/H$]${\sc  i}& 100 &  $-$0.027	&   +0.041 &  +0.023 & $-$0.044 & 0.016   &0.058    &0.024	 \\
$[$Fe/H$]${\sc ii}&  13 &  $-$0.119	&   +0.122 &  +0.040 & $-$0.030 & 0.045   &0.072    &0.055	 \\

\hline
\end{tabular}
\label{t:sensitivityuUNIFIN}
\end{table*}

\begin{table*}
\centering
\caption[]{Sensitivities of abundance ratios to variations in the atmospheric
parameters and to errors in the equivalent widths, and errors in abundances for
stars of NGC~6388 observed with GIRAFFE.}
\begin{tabular}{lrrrrrrrr}
\hline
Element     & Average   & T$_{\rm eff}$ & $\log g$ & [A/H]   & $v_t$    & EWs     & Total   & Total      \\
            & n. lines  &      (K)      &  (dex)   & (dex)   &kms$^{-1}$& (dex)   &Internal & Systematic \\
\hline        
Variation&              &  50           &   0.20   &  0.10   &  0.10    &         &         &            \\
Internal &              &   6           &   0.04   &  0.04   &  0.22    & 0.03    &         &            \\
Systematic&             &  57           &   0.06   &  0.02   &  0.01    &         &         &            \\
\hline
$[$Fe/H$]${\sc  i}& 20 &    +0.014     &   +0.032 &  +0.022 & $-$0.060 & 0.027   &0.067    &0.020	\\
$[$Fe/H$]${\sc ii}&  3 &  $-$0.079     &   +0.118 &  +0.040 & $-$0.018 & 0.070   &0.079    &0.097	\\
									   
\hline
\end{tabular}
\label{t:sensitivitymUNIFIN}
\end{table*}

\section{Results and comparison with other studies}

\begin{figure}
\centering
\includegraphics[scale=0.40]{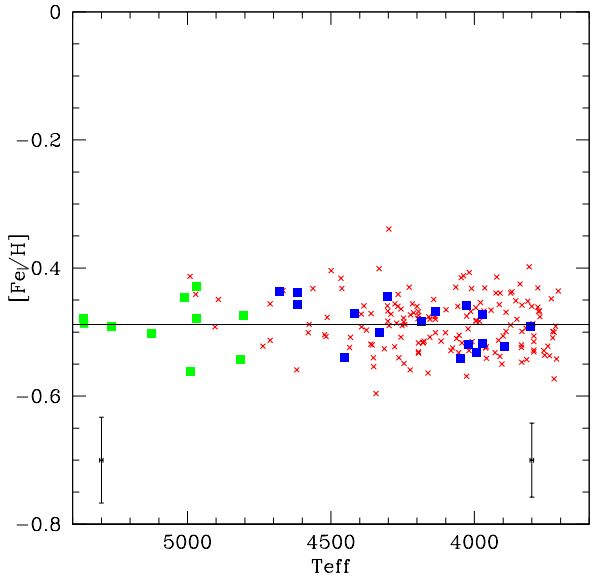}
\caption{Iron abundances in NGC~6388 as a function of effective temperature from
the present work and Carretta and Bragaglia (2018). Symbols for the evolutionary
stages are as in Fig.~\ref{f:cmdVKK}. Error bars on the left and on the right
are internal errors for abundances derived from GIRAFFE and UVES spectra,
respectively.}
\label{f:feteffnon}
\end{figure}

The first result of the present work is that the metallicity of NGC~6388 is
[Fe/H]$=-0.509\pm 0.011\pm 0.024$ dex ($rms=0.039$ dex, 12 stars) on our
metallicity scale defined by the high  resolution UVES spectra (Carretta et al.
2009c).
This result is strongly corroborated by the metal abundance derived from the
very large sample of stars with GIRAFFE HR13 spectra: 
[Fe/H]$=-0.488\pm 0.003\pm 0.020$ dex ($rms=0.040$ dex, 150 stars), where the
first and second term refer to statistical and systematic errors, respectively.
We plot in Fig.~\ref{f:feteffnon} iron abundances in NGC~6388 as a function of
the effective temperatures for the present sample, to which we added the stars
analysed in Carretta and Bragaglia (2018).

It  should  be  noted  that  the  observed  star-to-star scatter in iron
abundance is actually smaller than the estimate of the random errors (see 
previous Section). Hence the present data  do not support the existence of a
metal abundance spread in NGC~6388. This is the first conclusion from our
homogeneous and ample data set and will be further discussed below.

Abundances from Fe~{\sc ii} are in good agreement with results from neutral
transitions. We found [Fe/H]$=-0.491$ dex, $rms=0.039$ dex from the 12 stars
with UVES spectra, and [Fe/H]$=-0.478$ dex, $rms=0.053$ dex from 149 stars with
GIRAFFE spectra. Again, the estimates of the internal errors are consistent with
no significant intrinsic spread in metallicity.

Finally, we can combine the UVES sample of the present work to the sample in
Carretta and Bragaglia (2018), analysed exactly with the same procedure.
We obtain an average metallicity of [Fe/H]$=-0.480$ dex, $rms=0.045$ for the
total sample of 35 stars with UVES spectra, in excellent agreement with the
results from GIRAFFE spectra.

\begin{figure}
\centering
\includegraphics[scale=0.40]{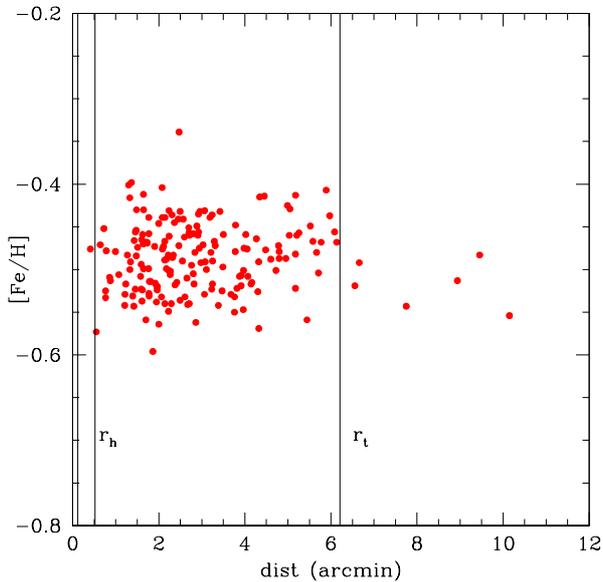}
\caption{Distribution of [Fe/H] ratios for our total sample (35 UVES and 150
GIRAFFE) as a function of the distance from the cluster center. The solid lines
are traced at the core radius, the half-mass radius, and the tidal radius
from Harris (1996).}
\label{f:distrFe}
\end{figure}

In Fig.~\ref{f:distrFe} we show the distribution of [Fe/H] values for our total
sample of 35+150 stars in NGC~6388 as a function of the radial distance from the
cluster centre. The bulk of our sample is enclosed between the half-mass radius
($r_h=0.52$ arcmin, Harris 1996) and the tidal radius ($r_t=6.21$ arcmin). Only
six stars, members according to their RV, are found beyond this limit. However,
it is not automatic to call them extra-tidal, as the value for $r_t$ is not
univocal and may be larger. In fact, Dalessandro et al. (2008) give 7.57 arcmin (29 pc) and 
Baumgardt et al. (2018) have a value about 3 times larger (based on a tidal
radius of 85.6 pc; the very large difference comes from the different methods in
estimating the cluster profile, since they also use dynamical information in
their modelling).
We identified five of the six stars with the Vasiliev \& Baumgardt (2021)
catalog of membership in Galactic GCs based on Gaia EDR3 and only one, the
closest to the
centre, is considered member. Currently, the Gaia astrometry has not yet reached
its full potentiality, especially in crowded fields as those in GCs. On the
other hand, RVs and metallicities for these five stars are indistinguishable from
the bulk cluster values, hence we consider these stars as cluster members, with
the cautionary flag that they may deserve further investigation.
However, their membership issue does not influence the main result of our work.

\begin{figure}
\centering
\includegraphics[bb=18 180 420 730, clip, scale=0.9]{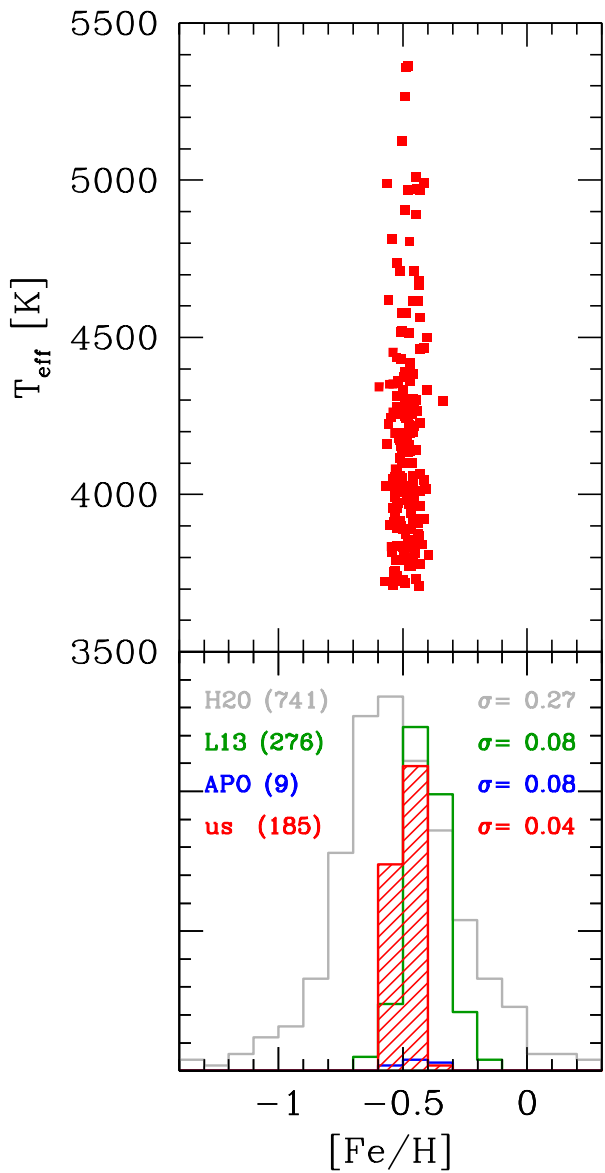}
\caption{Upper panel: distribution of our [Fe/H] values in NGC~6388 as a
function of T$_{\rm eff}$. Lower panel: histogram of metallicity in the various
studies considered: this work (red shaded histogram), L13 (green), M20 (APOGEE,
blue), and H20 (grey). For each sample the number of stars and the rms scatter
are also listed (see also Table~\ref{t:samples}).}
\label{f:istomet}
\end{figure}

We show in Fig.~\ref{f:istomet} (upper panel) the narrow [Fe/H] distribution in our sample, 
which displays no trend with T$_{eff}$. The lower panel shows the histograms of the [Fe/H]
values in our work and L13, M20, and H20, for comparison.
We also indicate the number of stars considered in the average and the rms
scatter obtained in each analysis. The information is summarised in
Table~\ref{t:samples} where we report also the type of spectra used in each
analysis and their spectral resolution. Further details are discussed in the
following sub-sections.

\begin{table}
\centering
\caption{Information on metallicity of different data sets.}
\setlength{\tabcolsep}{1.2mm}
\begin{tabular}{lllrrl}
\hline\hline
Sample & [Fe/H] & rms   &  Nr  & Resol & Note\\
\hline
us     & $-$0.488 & 0.040 &  150 & 26400 & GIRAFFE HR13 \\
       & $-$0.480 & 0.045 &   35 & 47000 & UVES \\
L13    & $-$0.401 & 0.078 &  276 & 18000 & GIRAFFE HR13,HR21 \\
M20    & $-$0.407 & 0.164 &   24 & 22500 & APOGEE,all stars\\
       & $-$0.436 & 0.077 &    9 &       & APOGEE, SNR$>$70 \\
H20    & $-$0.43  & 0.48  & 4098 &  2800 & MUSE; their tab.7 \\
       & $-$0.52  & 0.27  &  741 &       & HB method, see text \\
\hline
\end{tabular}
\label{t:samples}
\end{table}

\begin{figure}
\centering
\includegraphics[scale=0.40]{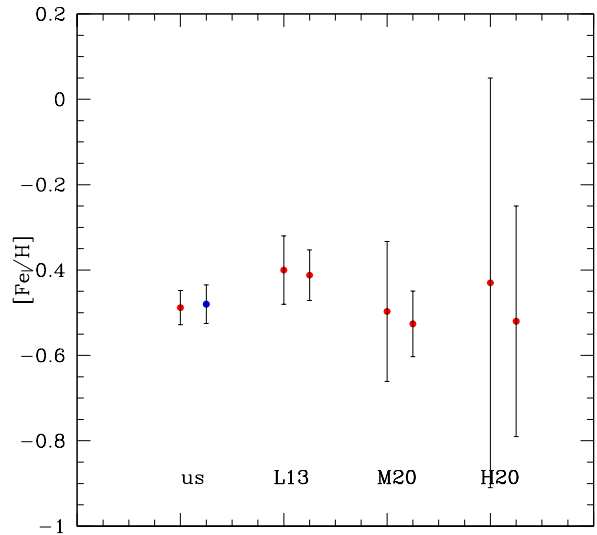}
\caption{Average metallicity and associated rms scatter for NGC~6388 resulting
from the present work (the blue circle on the right indicates the value from UVES spectra, the red
one is for GIRAFFE spectra) and from literature studies. The rightward point for
L13 refers to the analysis made with EWs (and not with the CaT). The rightward
point for M20 refers to the value obtained considering only the high S/N spectra
and the rightward point for H20 is derived from their HB method (see text).}
\label{f:meanfe}
\end{figure}

Mean [Fe/H] values and associated rms scatters from the present work and from
literature studies are plotted for NGC~6388 in Fig.~\ref{f:meanfe}. For the
present work we distinguish the metallicity obtained from UVES (right blue point) and
GIRAFFE (left red point) spectra. For the analysis by M20, the [Fe/H] value and the 
scatter on the right are from the selected sample of stars with S/N$>70$ (see
below). For H20, the rightward value is the one derived from their HB method
(see below).

\subsection{Comparison with L13}

Lanzoni et al. (2013) were mostly concerned with the kinematics of the cluster,
for which they acquired SINFONI low resolution, near-IR spectra in the innermost
part and FLAMES spectra (GIRAFFE setup HR21, plus UVES 580) in the external
regions, adding also FLAMES archive data.  
However, to better cull out non-member stars, they also derived metallicities
for a limited number of giants in their sample. The classical abundance analysis
with EWs was used for spectra available in the ESO archive at the epoch (UVES
and GIRAFFE HR11, HR13), whereas for the larger number of stars with HR21
spectra, metal abundances were obtained from a calibration of the CaT method. The
average value of [Fe/H] from 276 stars is $=-0.401$ dex, with $\sigma=0.078$
dex. Unfortunately, not many details nor an error budget were provided
concerning the abundance analysis, so that a formal conclusion about the
metallicity spread is not possible. However, we note that L13 did not claim that
an intrinsic spread was present in NGC~6388.

\begin{figure}
\centering
\includegraphics[scale=0.40]{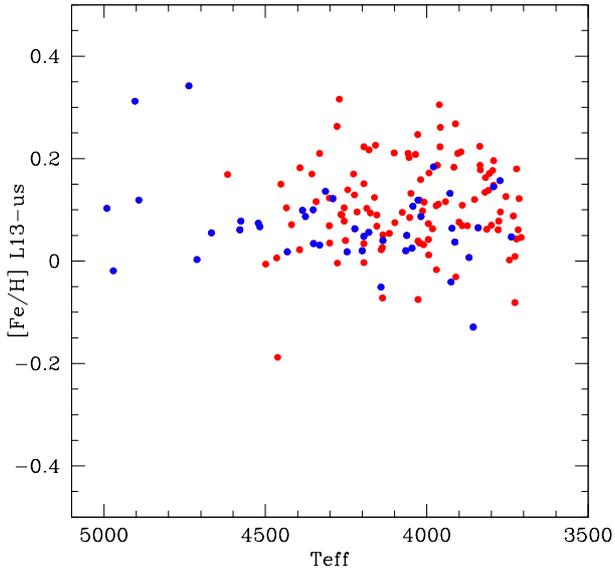}
\caption{Difference in metallicity between the present work and L13 as a
function of the effective temperature. Blue circles indicate stars whose
analysis in L13 was  based on EW measurement on UVES and GIRAFFE HR11, HR13
spectra. Red points refer to stars analysed with the CaT method from GIRAFFE
HR21 spectra in L13.}
\label{f:fenoil13}
\end{figure}

In Fig.~\ref{f:fenoil13} we plot the difference in [Fe/H] (in the sense
L13-us) as a function of effective temperature for 151 stars present in both
studies. The mean difference is 0.098 dex, with an rms scatter 0.086 dex.
Although scarcely significant, this offset can be ascribed to a number of
reasons, including a difference in the solar reference abundance, not provided
in L13.

However, we were able to identify 43 stars that were analysed by L13 using
methods similar to those adopted by us, with temperature and gravity from 
photometry and EWs measured on high or intermediate resolution UVES and GIRAFFE
HR11, HR13 spectra. Those are stars whose spectra were in the ESO archive in
2013, essentially those analysed in Carretta et al. (2007a, 2009b) plus a few
stars from the program by Catelan, recently studied by Carretta and  Bragaglia
(2018). We indicate them with blue symbols in
Fig.~\ref{f:fenoil13} and we can immediately see that the set of these stars
defines a sequence with a lower scatter with respect to the whole sample, apart
from three outliers. We found an average difference in [Fe/H] = 0.064 dex
(L13-us), with $\sigma=0.053$ dex from 40 stars, which have a mean metallicity
of -0.412 dex in L13, with $\sigma=0.059$ dex. This scatter is in good agreement
with our findings, it leaves no space for an intrinsic spread in metal
abundance, and it suggests that large part of the difference and scatter shown
in Fig.~\ref{f:fenoil13} has to be ascribed to stars whose metallicity was
obtained from HR21 GIRAFFE spectra using the CaT method.

\subsection{Comparison with M20}

M\'esz\'aros et al. (2020) published results based on APOGEE DR16 for 31 GCs;
for NGC~6388 there are 24 stars observed, but only 9 stars (of which 7 in common
with us) pass their suggested quality cuts, i.e., S/N$>70$ and T$_{\rm
eff}<5500$ K. Average values for them
are given in Table~\ref{t:samples}. In particular, for the high-quality
sample, the average value is [Fe/H]$=-0.436$ dex with a spread of 0.077 dex, to
be compared to an internal error of 0.152 dex (M20). Again, no indication of iron
spread is present. We recall that M20 detected a significant iron spread only
for $\omega$~Cen; they did not find metallicity variations for any of the
iron-complex GCs (type-II GCs) in their sample, i.e. NGC~362, NGC~7089,
NGC~6388 and, notably, NGC~1851 and M~22 (NGC~6656).

\begin{figure}
\centering
\includegraphics[scale=0.40]{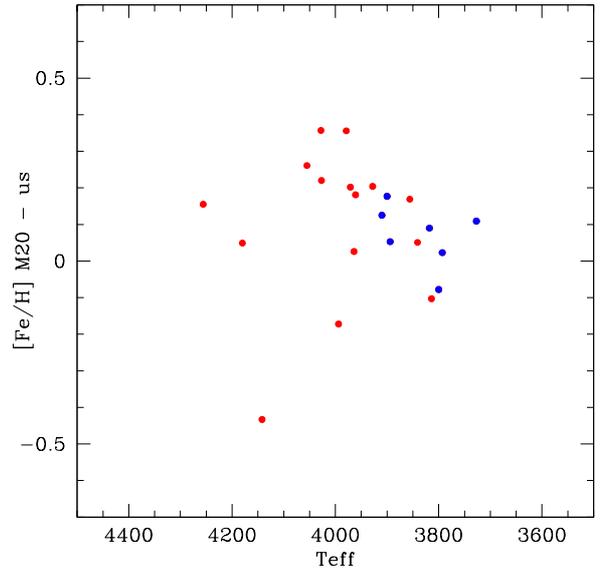}
\caption{Difference in metallicity between the present work and M20 as a
function of the effective temperature for stars in common between the two
studies. Blue circles indicate stars which pass in APOGEE the quality cuts
recommended by M20, namely S/N$>70$ and T$_{\rm eff}<5500$ K.}
\label{f:fenoiapo}
\end{figure}

For NGC~6388 we found 22 stars in common between the present work and M20. The
difference in [Fe/H] (in the sense M20 - us) is plotted in Fig.~\ref{f:fenoiapo}
as a function of the effective temperature. The average difference is 0.092 dex
($\sigma=0.178$ dex, 22 stars). On the other hand, if we consider only stars
that pass the quality criteria recommended by M20 (indicated by blue symbols in
Fig.~\ref{f:fenoiapo}), the difference is almost unaltered (0.072 dex), but the
dispersion is more than halved ($\sigma=0.082$ dex, 7 stars).
This comparison suggests that using lower S/N spectra may result into an
artificially larger spread, despite using the same method of abundance analysis. 
The average metallicity derived from the seven  stars in common, with S/N$>70$
spectra in APOGEE, is [Fe/H]$=-0.427$ dex. Taking into account the -0.09
dex offset due to different solar reference abundances, the value corresponds to
[Fe/H]$=-0.517$ dex on our abundance scale, in excellent agreement with the
results obtained in the present work. The points for M20 in Fig.~\ref{f:meanfe}
were also shifted to our scale of solar abundance by using the above offset.

\subsection{Comparison with H20}

The case of H20 is intricate, as they found a complex structure in the
metallicity distribution and a large dispersion (even larger than for
$\omega$~Cen, before taking into account the errors, see below). They employed
MUSE on RGB stars in 25 MW GCs, among which NGC~6388, to derive their
metallicity. The low-resolution MUSE spectra were acquired in the central parts,
i.e. the same covered by the HST {\em UV Survey} (see below), and were used to derive
metallicity through the CaT method. This technique requires a calculation of
$W'$, the so-called reduced EWs, which are then used to calibrate a relation
with metallicity, using literature reference values. Husser et al. (2020) used
four different methods to calculate $W'$, which they indicated as 'HB' (using
only stars brighter than V(HB)+0.2 and a linear relation with V-V(HB); this is
generally used in literature), 'all' (using all RGB stars and a quadratic
relation with V-V(HB)), 'M' (same as 'all', but using the absolute magnitude in
the F606W band), and 'lum' (same as previous ones, but using the luminosity),
see H20 for more details.

For NGC~6388, H20 cite an average value [Fe/H]$=-0.43~(\sigma=0.48)$ dex for the
whole cluster (4098 stars), and we calculated $-0.52~(\sigma=0.27$) for the 741
stars for which $W'$ is calculated with the HB method. As they were
studying the GCs in the multiple populations scenario, they also calculated the
metallicity separately for what they indicate as population P1, P2 (P and I+E in the
Carretta et al. 2009b division) and P3, which is the ``anomalous" component, on
the red RGB (see Milone et al. 2017 and below). The values for mean [Fe/H] and
$\sigma$ are -0.45, 0.45 dex for 579 P1 stars, -0.44, 0.42 dex for 1203 P2
stars, and -0.25, 0.39 dex for 411 P3 stars (note that these numbers sum up to
2193, i.e., not all the sample was assigned to a given component). We caution,
however, that the fraction of the P3 component in H20 (411/2193=0.187) is quite
different from the estimate by Milone et al. (2017) for NGC~6388 (0.299). Husser
et al. also claim to find indications of metallicity variations among P1 stars
for NGC~6388 and the other type-II GCs observed, but they do not consider it
definitely proved.  

The large dispersion and the difference in metallicity between P1,P2 and P3 are taken
with caution by H20, who comment that the errors are large for NGC~6388 and its
so-called twin NGC~6441, due to a combination of crowding affecting the
photometry, differential reddening, and high metallicity. For comparison,
$\omega$~Cen, with a well documented wide metallicity dispersion and a much
clearer structure and separation both in the CMDs and in the chromosome map (see
next section), has a similar dispersion, but with a smaller
associated error.  From Fig.~17 in H20, the uncertainty in [Fe/H] has a peak
value of more than 0.15 dex and a very broad, asymmetric distribution for
NGC~6388, compared to about 0.10 dex or less and a more peaked distribution for
most of the other GCs. In fact, once taken into account the uncertainties, the
intrinsic metallicity distribution for $\omega$~Cen has a $\sigma$ of about 0.4
dex (from their Fig.~18), while for NGC~6388 we read a value slightly more than
0.2 dex. At variance with the other studies, a real spread seems present,
according to H20.
However, we note that, due to the problematic photometry of this heavily
contaminated bulge cluster, affecting spectra extraction from MUSE data cubes, 
Pfeffer et al. (2021) concluded that NGC~6388 has no spread in iron.

\begin{figure}
\centering
\includegraphics[scale=0.40]{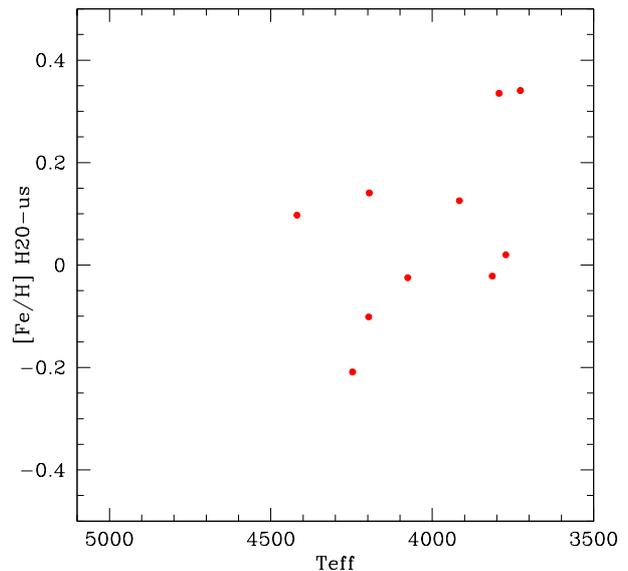}
\caption{Difference in metallicity between the present work and H20 as a
function of the effective temperature for stars in common. For H20,
metallicities obtained with the HB method were adopted.}
\label{f:fenoih20}
\end{figure}

Since the MUSE observations cover only the very central part of the cluster, the
intersection with our FLAMES data is small. We found only 11 stars in
common (10, considering the HB method) and we compared the metallicity results
in Fig.~\ref{f:fenoih20}. Since H20 only published EWs and corresponding
metallicitis, we used our values of T$_{\rm eff}$ in the figure, as in previous
comparisons with L13 and M20. The
mean difference in metallicity between our values and H20 is 0.070 dex (in the
sense H20 minus us), with $\sigma=0.176$, but a clear
trends is visible in the figure. At higher temperatures, the metallicities from
H20 are lower than ours, whereas at lower temperatures they find higher metal
abundances with respect to our values. Given the small number of stars in
common, we cannot pursue the comparison further and draw firm conclusions.

Finally, if we compare Fig.~20 in H20 with Fig.~27 in Marino et al. (2019), we
note a possible discrepancy in the arrow describing the effect of a change in
metallicity of 0.15 dex in the position in the ChM. In both cases an increase in
metallicity produces an increase in the $\Delta_{F275,F814}$ pseudocolour, but
while in the first case the arrow is directed upward, in the second it is almost
orthogonal and directed downward.

\subsection{No metallicity spread in NGC~6388}

The comparisons in the above sections point out the risks related to using low
resolution or low S/N spectra. In particular when dealing with metal-rich stars,
there is the danger to introduce a spuriously enhanced spread which is not an
intrinsic feature, but in part an artefact of the analysis.

On the other hand, the dispersion associated to the [Fe/H] values from our
present analysis, based on high resolution spectra, excludes the existence of a
noticeable intrinsic metallicity spread in the large sample of analysed stars in
NGC~6388, when coupled to the estimate of internal errors.
Two further tests were performed to confirm this result.

First, we used the algorithm illustrated by Mucciarelli et al.
(2012, kindly made available to us) to estimate the mean and intrinsic spread of
an elemental ratio ([Fe/H] in our case) by maximising the likelihood
function defined as in their Section 2.1. For NGC~6388 we obtained an average
value of -0.484 dex  and a zero intrinsic dispersion for the metallicity of the
185 stars in our  sample ($\sigma_{intr} = 0.000\pm 0.008$).

\begin{figure}
\centering
\includegraphics[bb=35 160 290 700, clip, scale=0.52]{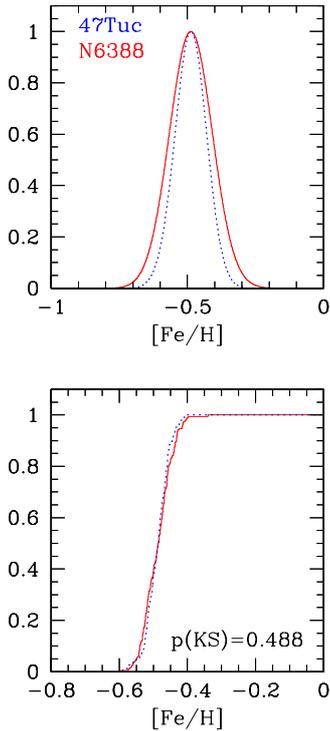}
\caption{Upper panel: generalised histogram of [Fe/H] values for NGC~6388 (red
solid line) and 47~Tuc (blue dotted line, from Carretta et al. 2009a,b),
normalised to their maximum. Lower panel: two-tail Kolmogorov-Smirnov test for
the above metallicity distributions. The K-S probability is also listed.}
\label{f:histogen}
\end{figure}

Furthermore,  we compare the generalised
histograms for NGC~6388 from the present work and for 47~Tuc from Carretta et
al. (2009a,b) in Fig.~\ref{f:histogen} (upper panel). The GC 47~Tuc was selected because it is analysed exactly in the same
way as NGC~6388,  its sample is almost as large (147 stars) as the present
sample, and it is the most metal-rich GC in our homogeneous FLAMES survey,
beside NGC~6388. Last but not least, 47~Tuc is one of the best representative of
the class of monometallic GCs, never suspected to host a  metallicity dispersion
(see Carretta et al. 2004, Carretta et al. 2009a,b, Cordero et al. 2014).
The sigma of the gaussian centered at each data point used to generalise the histogram was chosen
equal to the error on each point, namely 0.067 dex for NGC~6388 (from
Table~\ref{t:sensitivitymUNIFIN}) and 0.047 dex for 47~Tuc (from Carretta et al.
2009b). These values correspond to the internal errors derived for stars with
GIRAFFE spectra, representing the vast majority in the samples. 

After applying an offset of 0.255 dex to account for the difference in the mean
metallicity of the two GCs, the histograms are superimposed in the upper panel of Fig.~\ref{f:histogen}. 
The metallicity distribution for each cluster is well described by a single
symmetric Gaussian curve.
This comparison reveals only a negligible difference in the
spread, that can be fully ascribed to the different internal errors associated
to the analyses. A two-tail Kolmogorov-Smirnov test (lower panel of 
Fig.~\ref{f:histogen}) clearly indicates that the null hypothesis (the two
distribution are extracted from the same parent population) cannot be rejected.

Our analysis, the above tests, and the comparison with other literature data 
(see above) allow to conclude
that in NGC~6388 there is no evidence for an intrinsic metallicity dispersion.
This is an important information for a type-II GC. 

\section{Discussion}

The so called ``anomalous" RGB stars should be subjected to further 
investigation to clarify their nature and role in defining a separate class of
GCs. In the next Sections we will associate the metallicities we obtain from our
large sample to the UV HST photometry and we will discuss some properties of the
so called type-II GCs.

\subsection{Chromosome map and metallicity}

Neither Piotto et al. (2015) nor Milone et al. (2017) published a
classification of individual stars in their full  GC sample into FG, SG, and
red-RGB stars. These data only exist for six GCs out of 58 objects, and
unfortunately NGC~6388 (or any other type-II GC) is not among them.

We then downloaded the available HUGS catalogue (Nardiello et al. 
2018a)\footnote{https://archive.stsci.edu/prepds/hugs/}, who published the
photometry uncorrected for reddening and without distinction
between stars of different populations.
Starting from these data we followed the necessary steps to produce the
pseudo-colours maps used in that survey to single out different populations in
GCs. However, we did not apply corrections for differential reddening, which is
outside the goal of the present study and which is not indispensable to establish the lack of any metallicity variation among stellar populations in NGC~6388.

\begin{figure}
\centering
\includegraphics[bb=25 440 290 710,clip,scale=0.9]{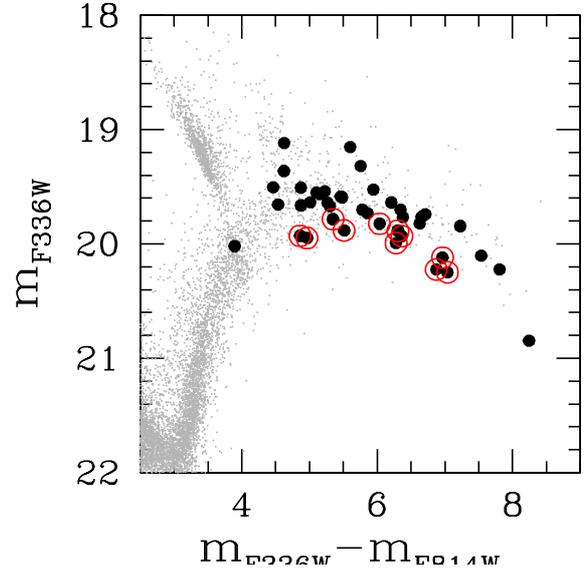}
\caption{Plot of $m_{F336W}-m_{F814w}$ versus$m_{F336W}$ for NGC~6388 (selecting
only stars with valid {\em sharpness} parameter (see text). Stars in our
spectroscopic sample are indicated by larger black dots, while stars selected on
the redder part of the RGB are shown as red circles. }
\label{f:ui}
\end{figure}

From the CMD shown in Fig.~\ref{f:ui} we identified stars in common between our
spectroscopic sample and the HST photometry by Nardiello et al (2018a). A few
stars along the redder part of the RGB have been highlighted (red circles) in
the plot to reproduce Fig.~16 in Milone et al. (2017). These red-RGB stars also
show up in a given region of the ChM only in type-II GCs.

\begin{figure*}
\centering
\includegraphics[scale=1.2]{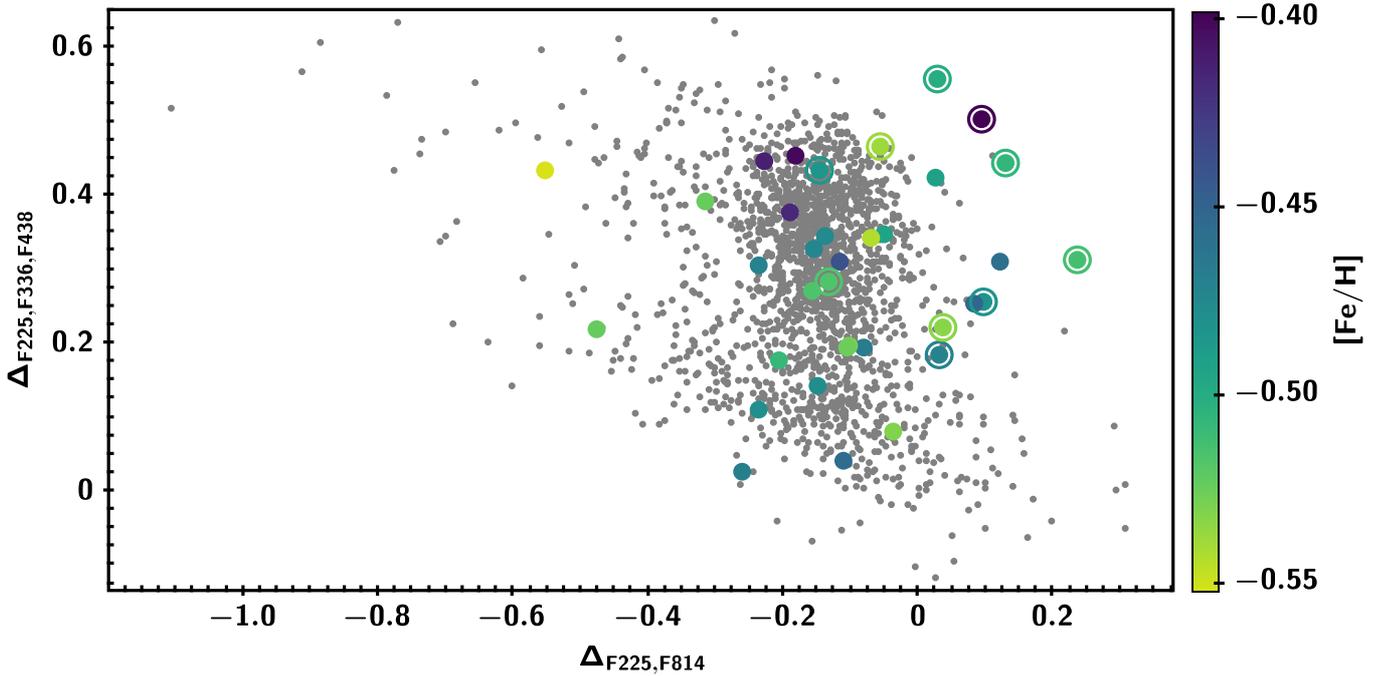}
\caption{Position of the stars in NGC~6388 in the 
so-called ChM map (Milone et al. 2017). Stars also in our spectroscopic
sample are showed as large  filled circles, coloured with their metallicity.
Stars selected on the red part of the RGB (see text and previous figure) are 
indicated by large open circles.}
\label{f:cromo}
\end{figure*}

Following the discussion in Nardiello et al. (2018a), we kept only stars with
good quality photometry. In particular we excluded stars with the {\em sharpness}
parameters outside the $\pm0.15$ range in any of the filters; no cut on error
was applied,  as all the stars involved in the ChM are giants and the
corresponding errors are very small and uniform. As in Milone et al. (2017) we
used stars in the magnitude range $m_{f814w}=14-18$, computed the colours and
pseudo-colours $col_1=(m_{F275w}-m_{F814w})$ and $col_2=(m_{F275w}-2\times
m_{F336w}+m_{fF38w})$, and defined lines describing the limits of the RGB stars
in the corresponding plots. Finally, we computed the values for
$\Delta_{275,814}$ and  $\Delta_{275,336,438}$   from eqs. 1, 2 in Milone et al.
and using the RGB widths in their Table~2). 
The resulting two-pseudocolours diagram is shown in Fig.~\ref{f:cromo}, which
reproduces sufficiently well the plot in Fig.~3 of Milone et al. (2017) and the
location of the different populations: FG stars, SG stars and the region where
the ``anomalous" red-RGB stars are arranged in this diagram.
Further refinement is outside the main goal of the paper and we await the
publication of the ChM by the group who has produced them 
homogeneously for further discussion.

In Fig.~\ref{f:cromo} we have identified the few stars selected on the redder
RGB with concentric open circles. 
Milone et al. (2017) claim that all evidence supports that stars in the red-RGB
are enhanced in C+N+O, in s-process elements, and in iron.
This is for instance true for NGC~1851, where we detected both a small iron
spread and an enhancement in s-process elements on the red-RGB we defined using
Str\"omgren filters (Carretta et al. 2010, 2011; see also below). However, we do
not find a significant spread in Fe in NGC~6388, and in particular we do not
find a metallicity enhancement for the red-RGB stars (see Fig.~\ref{f:cromo}). 
We stress that this conclusion is not strictly related to our method of
reproducing the ChM, which shows resemblance both to the one in M17 and to that
recomputed by H20.
Milone et al. (2017) list a fraction $0.299\pm 0.016$  of type-II stars with
respect to the total number of analysed stars in NGC~6388, while H20 have a
lower value. A third (or about one fifth, as in H20) of stars with enhanced
metal abundance would have been easily discerned in the distribution of [Fe/H]
values (considering that we observed stars also on the red-RGB), resulting in a
clearly detectable iron spread. From our data,  this is not the case.

\begin{figure}
\centering
\includegraphics[scale=0.40]{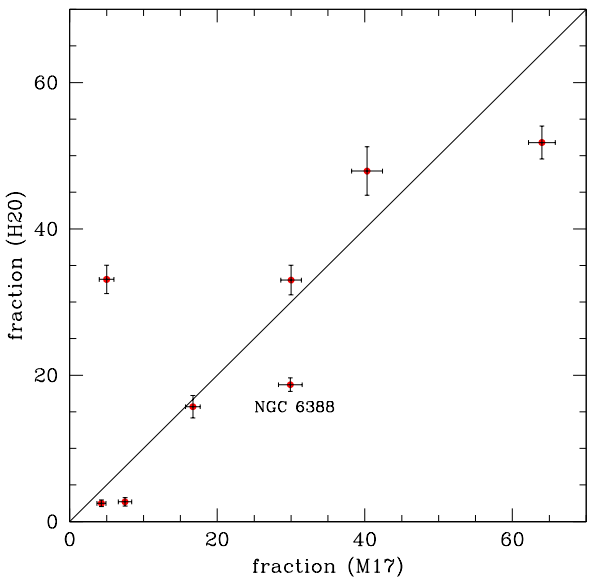}
\caption{Comparison of fractions of type-II stars in 8 GCs in common between the
studies by H20 and M17 (plus Nardiello et al. 2018b for NGC~7078/M~15). 
The line of equality is also shown and the position of NGC~6388 is labeled in the plot.}
\label{f:husserp3}
\end{figure}

\subsection{What really is a type-II cluster? }

With the present large sample of stars analysed spectroscopically, we can add
NGC~6388 as a good test to compare spectroscopy and UV photometry concerning
the classification of GCs.
When discussing the type-II GCs as a separate class, one must first note that
the fraction of 'peculiar' stars deduced from the ChM and photometric methods
has some degree of subjectivity, as evident from Table~\ref{t:fraction} and
Fig.~\ref{f:husserp3}, where we compare the fraction of the so called type-II
stars for eight GCs analysed by two groups starting from the same HST
photometry.
Values of the fractions and associated errors for M17 are taken from their
Table~2 (for NGC~1851 we use the fraction corrected for a misprint as in Gratton et al.
2019, their Table 5). To those, we added the corresponding value
for NGC~7078 (M~15) from Nardiello et al. (2018b) who considered also this GC as
a type-II cluster from the analysis of its pseudo-colour diagrams. For H20, we
computed the ratios using the fraction of their P3 stars over the total of
P1+P2+P3 populations from their Table~7. For the very complex cluster
$\omega$~Cen, we adopted as type-II stars the sum of the populations from P4 to
P9, as a visual comparison with M17 showed that the normal stars are only 
restricted to population P1+P2+P3.  The error bars are assumed from the Poisson
statistics.  

\begin{table}
\centering
\caption{Fraction of stars in type-II GCs from H20 and M17}
\begin{tabular}{rcc}
\hline\hline
GC     &  fraction        &   fraction \\
       &   H20            &     M17  \\
\hline
 362   & $0.027\pm 0.006$ & $0.075\pm 0.009$ \\
1851   & $0.330\pm 0.020$ & $0.300\pm 0.014^a$ \\
5139   & $0.518\pm 0.023$ & $0.640\pm 0.018$ \\
5286   & $0.157\pm 0.015$ & $0.167\pm 0.010$ \\
6388   & $0.187\pm 0.009$ & $0.299\pm 0.016$ \\
6656   & $0.479\pm 0.033$ & $0.403\pm 0.021$ \\
7078   & $0.331\pm 0.019$ & $0.050\pm 0.010^b$ \\
7089   & $0.025\pm 0.005$ & $0.043\pm 0.006$ \\
\hline
\end{tabular}
\begin{list}{}{}
\item[$^{a}$] from Gratton et al. (2019)
\item[$^{b}$] from Nardiello et al. (2018b)
\end{list}
\label{t:fraction}
\end{table}

Figure~\ref{f:husserp3} shows that the estimates of the fractions are more or
less distributed around the equality line, but in a few cases the values can be
very different, although the underlying photometric database is the same.
The only GC for which there is a formal agreement within one Poisson error bar
is NGC~5286. In other cases, such as NGC~362, NGC~1851, and $\omega$~Cen, the
discrepancies are not large. On the other hand, NGC~7078 (M~15) and
NGC~6388  have the  largest differences in the estimated fractions.
This indicates that a more objective criterion to separate the so called
type-II stars is required.

Milone et al. (2017) state that a type-II GC is  identified by (at least) one of the 
following properties: split SGB even in optical photometric bands, multiple
sequences in the ChM, and wide range in heavy elements, including
iron and extending to the species produced by $s-$process.
The sample of so called type-II GCs seems to be, however, a heterogeneous class. 
This motley group contains the two most massive GCs of our Galaxy,
that most probably were the nuclear star clusters of past dwarf galaxies later
incorporated in the Milky Way, namely NGC~5139 ($\omega$~Cen) and NGC~6715 (M~54,
see e.g. Bekki and Freeman 2003, Bellazzini et al. 2008, Carretta et al. 2010b;
see also Massari et al. 2019 who associate the first to Gaia-Enceladus and
confirm the association with Sagittarius of the second, using Gaia data).  Both
GCs probably experienced multiple bursts of star formation over their entire
life (see e.g. Johnson and Pilachowski 2010, Siegel et al. 2007). In both cases,
the metal-rich component according to the spectroscopy represents a fraction of
the total mass roughly in agreement with the estimated fraction of type-II
stars from photometry, with evidence of enhancement also in $s-$process
elements.

However, to the same group also belong two smaller GCs, less luminous by two 
orders of magnitude: NGC~1261 and NGC~6934. Small spreads in [Fe/H] (0.1-0.2
dex) were detected for these lower mass GCs attributed to the type-II class,
(Mu\~noz et al. 2021, Marino et al. 2021 for NGC~1261 and Marino et al. 2021 for
NGC~6934). These results are  based on samples of limited size, much smaller
than for the GCs discussed above. Marino et al. (2021) also note that for both
clusters no enhancement in $s-$process elements is found and suggest this is due 
to the low mass of these GCs. A comparison of the fractions determined by
photometry and spectroscopy would be meaningless, since only 8-14 stars in
NGC~1261 and 13 stars in NGC~6934 were observed with high resolution
spectroscopy.

Coming to the intermediate-mass GCs, 
the existence of a secondary, redder RGB sequence in a few GCs (besides
$\omega$~Cen) is known since more than one decade (it was discovered in NGC~1851
by Han et al. 2009 for the first time, using the U and I filters). Carretta et
al. (2011, 2013) discussed the cases for  NGC~1851 and NGC~362 using Str\"omgren
photometry and their own spectroscopic results. 
In NGC~362 Carretta et al. (2013) discovered that in the Str\"omgren $v-y$
colour the RGB is clearly split, with a minority of stars ($\sim 6\%$) defining
a secondary, redder sequence only populated by Ba-rich stars (roughly corresponding in size
to the type-II component claimed from UV photometry in M17). The same phenomenon
was also traced by Carretta et al. (2011) in NGC~1851, where the redder $v-y$
colour was attributed most likely to large enhancement in N (see also Carretta
et al. 2014 and Villanova et al. 2010).

\begin{figure}
\centering
\includegraphics[scale=0.40]{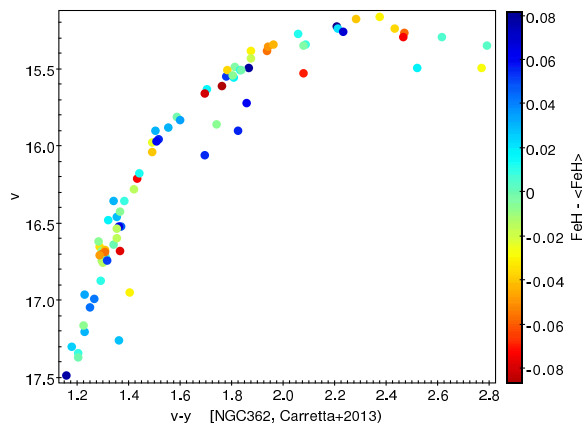}
\caption{Str\"omgren CMD $v-y$ vs $v$ from Carretta et al. (2013). Stars are
coloured with a palette indicating the deviations from the average metallicity
of the cluster. }
\label{f:n362vyv}
\end{figure}

Both GCs are now classified as type-II clusters. However, in NGC~362 no spread
in [Fe/H] was detected above the small level (0.041 dex) fully explained by
internal uncertainties in the abundance analysis. This is well shown in
Fig.~\ref{f:n362vyv} where the CMD of NGC~362 in Str\"omgren photometry is
coloured by the difference in iron with respect to the average [Fe/H] value in
this GC. However, the trickling of stars in this red sequence does not present
any enhanced [Fe/H] distribution with respect to the other giant stars.

In NGC~1851, instead,  a small metallicity dispersion was found; however, the
best results were obtained from a cluster analysis using a $k-$mean algorithm
and the run of [Fe/H] as a function of [Ba/H] as representative of the
$s-$process elements (Carretta et al. 2011). With this technique, two stellar
component were unveiled, the most metal-rich being also more rich in Ba and
$s-$process elements, with Str\"omgren photometric indices consistent with an
overabundance in carbon. The fraction of this component from spectroscopy is
roughly compatible with the fraction of type-II stars from the HST photometry,
both from M17 and H20.

The largest fraction of type-II stars, excluding $\omega$~Cen and M~54, belongs to
NGC~6656 (M~22), which has an iron spread of about 0.14 dex  according to Marino et
al.  (2011), besides all other characteristics of type-II clustes. As a caveat,
the metallicity spread has been  questioned by Mucciarelli et al. (2015)
who attributed the result to the adoption of surface gravities derived from
spectroscopy. Those values correspond to unrealistic low stellar  masses,
whereas using photometric $\log g$ values no evidence of spread was found from
singly ionised iron, at variance with Fe~{\sc i}. Mucciarelli et al. concluded
that this pattern could be explained by non-LTE effects, namely 
overionisation,
lowering abundances from Fe~{\sc i} while leaving unaltered those from 
Fe~{\sc ii}.
The surface gravity values used in our present analysis for NGC~6388 (the fourth
in line for type-II stars fraction, according to M17) were derived from
photometry and we did not find evidence of different iron abundances in the
large fraction of stars dubbed as type-II, as discussed above.

To conclude, in a fraction of GCs classified as type-II no metallicity spread 
is detected over large sample of stars. Some GCs show an enhancement in
$s-$process elements but not in iron (e.g. NGC~362). In other cases, such as
NGC~1261 and NGC~6934, the contrary is observed. Finally, for M~15 the
large dispersion measured for neutron-capture elements is known to be originated
by the $r-$process (e.g. Sneden et al. 2000, Sobeck et al. 2011, Worley et al.
2013, Kirby et al. 2020) and not by the the $s-$process, as found for some
other  type-II GCs. Models of fast rotating massive stars (also called
``spinstar", Frischknecht et al. 2016, Limongi and Chieffi 2018) would show an
increase of $s-$process yields, but would provide also alterations in
light-elements, that are not seen correlated to abundance variations in
neutron-capture elements in M~15 or other GCs (e.g. Roederer 2011).
Moreover, no evidence of a spread in iron is 
claimed on the basis of extant high resolution spectroscopy (e.g. Carretta et
al. 2009a,b).

All the above considerations were based on the metallicity spreads (or their
absence) derived in the original studies. For a more homogeneous approach, we
evaluated the iron dispersion of each type-II GC using the same method (see
Sect. 4.4) we employed for NGC~6388, i.e. by maximization of the likelihood 
function as defined by Mucciarelli et al. (2012), that estimates the intrinsic
spread by taking into account also the errors.
In addition to the GCs listed as type-II in M17 and Nardiello et al. (2018b), 
we also added NGC~6273, as done in Gratton et al. (2019). This GC was not 
classified by M17 as a type-II because it lacks the UV HST photometry, however it 
clearly shows a large iron variation among its stellar populations (Johnson et
al. 2015, 2017). Results are given in Table~\ref{t:intri}, where we list the 
adopted internal error for the metallicities of the individual stars, the
derived intrinsic  spread and the associated uncertainty, the number of stars,
and the reference for the abundance analysis.
To keep on the safe side, we conservatively chose to use the internal errors
derived for stars with GIRAFFE spectra (usually larger than those for stars with
UVES spectra) for NGC~6388 and the other GCs from our group (NGC~362, NGC~1851,
NGC~6715, NGC~7078).

\begin{table}
\centering
\caption{Intrinsic spreads in iron}
\setlength{\tabcolsep}{1mm}
\begin{tabular}{llllrl}
\hline
\hline
NGC & error &$\sigma_{intr}$ & $err_\sigma$ &stars & Ref\\
\hline
362  & 0.041 & 0.000 &0.014 & 92 & Carretta et al. 2013\\
1261 & 0.060 & 0.098 &0.034 &  8 & Mu\~noz et al. 2021\\
         & 0.090 & 0.000 &0.030 & 14 & Marino et al. 2021\\
1851 & 0.043 & 0.026 &0.006 &124 & Carretta et al. 2011\\
5139 & 0.117 & 0.247 &0.007 &855 & Johnson \& Pilachowski 2010\\
5286 & 0.075 & 0.078 &0.013 & 62 & Marino et al. 2015 \\
6273 & 0.026 & 0.181 &0.018 & 51 & Johnson et al. 2017 \\
6388 & 0.067 & 0.000 &0.008 &185 & this paper \\
6656 & 0.100 & 0.000 &0.112 & 35 & Marino et al. 2011\\
6715 & 0.026 & 0.184 &0.015 & 76 & Carretta et al. 2010c \\
6934 & 0.090 & 0.089 &0.035 & 13 & Marino et al. 2021\\
7078 & 0.040 & 0.042 &0.008 & 52 & Carretta et al. 2009a,b\\
7089 & 0.072 & 0.268 &0.054 & 14 & Yong et al. 2014 \\
\hline
\end{tabular}
\label{t:intri}
\end{table}

From this Table we see that the procedure confirms the zero intrinsic 
metallicity spread for NGC~362, and the large iron spreads in NGC~5139, 
NGC~6273, NGC~6715, and NGC~7089. Smaller but robust spreads are derived in 
NGC~6286 and NGC~6934. A few GCs merit further comments. The case
of NGC~1851 was already discussed above: the best separation of populations
requires both [Fe/H] and the [Ba/H] abundances,  but a small spread is still
confirmed by iron abundances only. 
 
The case of NGC~1261 is more uncertain. For this GC, two studies, both resting 
on small samples, give different results, although in both the original papers
the quoted spread is the same (about 0.1 dex). The cause is probably related to
the smaller internal error estimated in Mu\~noz et al. (2021), whereas the one
in Marino et al. (0.09 dex) is almost the same of the observed iron
spread.

The null spread for NGC~6656 is {\em prima facie} surprising. M~22 seems to be 
a genuine type-II GC, showing all the required features; however, the rather
large internal error from Marino et al. (2011), that we derived from an average
of all values associated to spectra from different telescopes, may reconcile 
the present finding with the usually assumed value of about 0.15 dex spread in
the cluster. 
Finally, with this method also NGC~7078 seems to have a (small) intrinsic 
iron spread, contrarily to what derived from a simple comparison of observed
spread and internal errors in the analysis. However, M~15 is the most metal-poor
GC in our FLAMES survey, and the analysis of GIRAFFE spectra was rather
difficult. Conservatively, for the value derived in Table~\ref{t:intri} we
retained only stars with at least six measured Fe lines.
This cluster surely deserves merits further attention and a larger data set of
good quality spectra. A paper on a new abundance analysis in M~15 is in
preparation. 

\begin{figure*}
\centering
\includegraphics[bb=20 440 580 660, clip, scale=0.80]{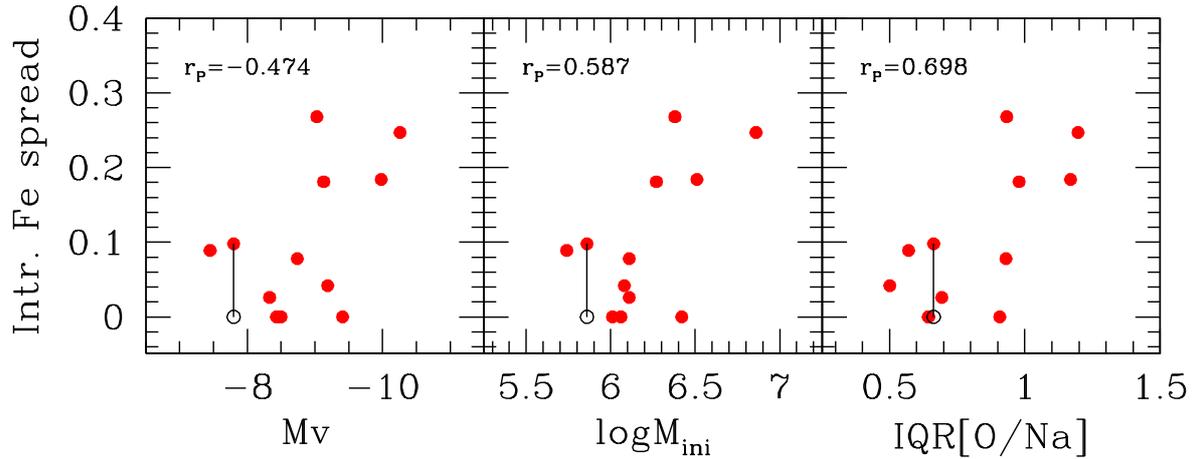}
\caption{Intrinsic spread in iron as a function of the total cluster absolute
magnitude $M_v$ (from Harris 1996, 2010 edition), logarithm of the initial mass
(from Baumgardt et al. 2019), and IQR[O/Na] (from Carretta 2019), from left to
right. In each panel we indicate the Pearson coefficient $r_P$ (see text for the
significance level). A line connects the two values for NGC~1261 from different
studies.}
\label{f:mlout}
\end{figure*}

The intrinsic iron spread does not show a significant correlation with the GC
present-day masses (as represented by the model-independent total absolute
luminosity $M_V$), as shown in the left panel of Fig.~\ref{f:mlout}. The 
probability associated to the Pearson linear regression
coefficient is only $p=0.119$. We thus confirm the findings by Mu\~noz et 
al. (2021) who found no trend with the mass, although using a more heterogeneous
set of observed spreads. In Fig.~\ref{f:mlout}, a solid line connects the two
different values derived for NGC~1261, but all the regressions are computed using
the non-null spread for this GC from Mu\~noz et al. (2021). A marginally
significant ($p=0.045$) trend is observed as a function of the initial total
cluster mass (central panel of Fig.~\ref{f:mlout}, estimated by Baumgardt et al.
(2018). This correlation is probably due to the fraction of SN ejecta retained
in type-II GCs being a function of the (original) cluster mass, as discussed
e.g. by Renzini et al. (2015) and Gratton et al. (2019).

Finally, in the right panel of Fig.~\ref{f:mlout} we plot a relation with a high
level of significance ($p=0.012$) showing that the intrinsic spread is well
correlated to  IQR[O/Na], the interquartile range of the [O/Na] abundance ratio,
from Carretta (2019). This quantity is a measure of the extent of the Na-O
anti-correlation. It may seem odd to see this quantity connected to the iron
spread in GCs, as none of the early producers of the alterations in light
elements is predicted to pollute the cluster environment with additional iron.
However, we remind that each of the involved quantities is a strong function of
the GC total mass (see e.g. Carretta 2006, 2019, M17) and this occurrence
well explains the above correlation.

To conclude this discussion, the evidence suggests that either the group 
of GCs called type-II is  heterogeneous or NGC~6388 is not a genuine type-II GC.
Together with previous analysis, the present results for NGC~6388 would require
to explain the origin of the  red-RGB sequence and their position in the ChM, as
well as the discrepant fractions found for some GCs by different groups,
which claims for a less subjective definition or more published data.

\section{Summary}

We present a large sample of homogeneous metallicities derived for giant stars
in the massive bulge GC NGC~6388. We combined new observations
with archival data and previously published data obtained with intermediate and
high resolution spectroscopy for a total of 150 stars with spectra from GIRAFFE
(high resolution HR13 setup) and 35 stars with spectra from UVES/FLAMES at the 
ESO-VLT UT2 telescope. 
Radial velocities, atmospheric parameters and resulting abundances of [Fe/H]
were homogeneously obtained with the same methodology and on the same scale used
for the other GCs in our FLAMES survey (e.g. Carretta et al. 2006, 2010a). 

The most important result we found is that NGC~6388 does not show any intrinsic
metallicity dispersion, at odds with what found by H20 using low-resolution
MUSE spectra. The $rms$ values obtained from the 35 stars with UVES
spectra and 150 stars with GIRAFFE spectra are 0.045 dex and 0.040 dex, respectively, fully
compatible with the uncertainties of the analysis. 
Our result confirms the absence of a significant metallicity spread found by all
other previous works based on spectra with sufficiently high resolution and S/N:
Carretta et al. 2007a, Carretta et al. 2009b, Carretta  \& Bragaglia 2018, L13
from GIRAFFE spectra; M\'esz\'aros et al. 2020 from APOGEE  spectra). 

We discussed our findings also in connection with the ChM of NGC~6388 and other
type-II GCs. About a third of giant stars in NGC~6388 is supposed to be located
in the region populated by red-RGB, ``anomalous" stars defining the nature of
type-II GCs according to the UV pseudo-colour diagram. However, no difference in
metallicity is noticeable among these stars in NGC~6388 from our large sample
studied with high resolution spectra. We conclude that either an intrinsic metallicity
dispersion is not a necessary requisite for type-II GCs or NGC~6388 does not
belong to this group. In the latter case, it would be necessary to understand 
the origin of the red-RGB stars on the ChM for this GC, as well as for M~15 and
NGC~362 where an intrinsic metallicity dispersion is not found from
spectroscopic analysis of large samples of stars.

The homogeneous data presented in this paper are also the necessary first step
for a complete characterisation of the chemistry of multiple stellar populations
in NGC~6388, their relation with the cluster dynamics (e.g. rotation in the
individual sub-populations), and its use to better define the origin of this
massive GC in the context of the GC population of the Milky Way. All these
issues will be explored in detail in a forthcoming paper.

\begin{acknowledgements}
We made extensive use of the ESO archive and we thank the personnel maintaining
it and developing and applying the instrument pipelines. This research made use
of observations made with the NASA/ESA Hubble Space Telescope obtained from the
Space Telescope Science Institute, which is operated by the Association of
Universities for Research in Astronomy, Inc. under NASA contract NAS 5-26555 We
warmly thank Raffaele Gratton for useful discussion and Michele Bellazzini for
sharing his code to measure intrinsic spreads.  This research made use of the
SIMBAD database (in particular Vizier), operated at CDS, Strasbourg, France, of
the NASA Astrophysical Data System, of IRAF, and of TOPCAT
(http://www.starlink.ac.uk/topcat/, Taylor 2005).  
\end{acknowledgements}

\end{document}